# Study of the asymptotic motion of a sporting projectile taking into account the Magnus force

Peter Chudinov

*Department of Engineering, Perm State Agro-Technological University, 614990, Perm, Russia*

**Abstract.** A classic problem of the motion of a projectile thrown at an angle to the horizon is studied. Air resistance force and Magnus force are taken into account with the use of the quadratic laws. We consider the asymptotic motion of the projectile, i.e., the motion on a sufficiently large time interval. The equations of motion are written both in Cartesian coordinates and in natural axes. The aim of the study is the analytical determination of the characteristics of asymptotic motion - the limiting angle of inclination of the trajectory to the horizontal and the terminal velocity, and obtaining an analytical representation for the velocity hodograph. No restrictions are imposed on the initial throwing conditions and other parameters. The equations of motion in Cartesian coordinates are used to find the angle of inclination of the trajectory and the terminal velocity. The equations of motion in natural axes are used to determine velocity hodograph. The velocity hodograph is defined in the form an approximate implicit analytical formula linking the trajectory angle of the projectile to its velocity. The motion of a golf ball, is presented as example. Numerical calculations show a complete coincidence of analytically found values of required quantities with numerically found values. The proposed analytical formulas can be useful for all researchers of this classical problem.

**Keywords**: Projectile motion; quadratic resistance law; Magnus force.

## 1. Introduction

The study of the motion of sports projectiles is an important part of ballistics. Such motion is studied mainly by numerical methods, especially when air resistance and the Magnus effect are taken into account [1-5]. Nevertheless, in some cases it is possible to analytically determine some characteristics of the motion. In [2], closed-form solutions for the object velocity are presented either when the quadratic drag force is negligible or when the quadratic Magnus force is negligible. In [2] the case when both of these forces act together but are very small is also considered. In [5], the motion of a projectile in near-vertical motion was analytically investigated. It is known that in the presence of a drag force, the trajectory of the projectile has an asymptote. The study of the asymptotic motion of the projectile taking into account the quadratic drag force and the Magnus force is the first objective of this study. Asymptotic motion is defined as motion at time intervals at which the trajectory of the projectile approaches an asymptote. No restrictions are imposed on the magnitudes of forces and initial conditions of the projectile. The second aim of the paper is to obtain an analytical equation of the velocity hodograph, albeit an approximate one.

Let us consider the motion of a projectile of mass $m$ thrown into air under the gravitational force $mg$ as depicted in Fig. 1. The gravitational force is supplemented by two more fundamental forces: the quadratic drag force $F_D$ due to air resistance and the quadratic Magnus force $F_L$ owing to the object's spinning lift. Let's assume that ehe projectile (ball) is spinning counterclockwise at a constant angular velocity $\omega$ around its axis parallel to the axis **z**. In this case, the Magnus force lies in the **Oxy** plane as shown in Figure 1. The drag force and Magnus force are defined according to [1, 2] as

$$F_D = mgk_1 V^2, \quad F_L = mgk_2 V^2. \tag{1}$$

Here $k_1, k_2$ - are the coefficients of proportionality of the drag force and Magnus force, respectively. They are determined by the relations

$$k_1 = \rho A C_D / (2mg), \quad k_2 = \rho A C_L / (2mg).$$

Here $\rho$ is the air density, $A$ is the cross-sectional area of the spherical projectile, $C_D$ is the (dimensionless) drag coefficient, $C_L$ is the (dimensionless) lift coefficient. For Reynolds number values $10^3 \leq Re \leq 3 \times 10^5$ coefficient $C_D$ is assumed to be constant and equal to 0.45. For sports projectiles (cricket, tennis and golf balls) we assume that $C_D = 0.45$. The coefficient $C_L$ is calculated according to the generally accepted formula [1]

$$C_L = 0.319 \left[1 - \exp\left(-2.48 \times 10^{-3} \omega\right)\right],$$

where the angular velocity $\omega$ of the projectile is measured in rad/s.

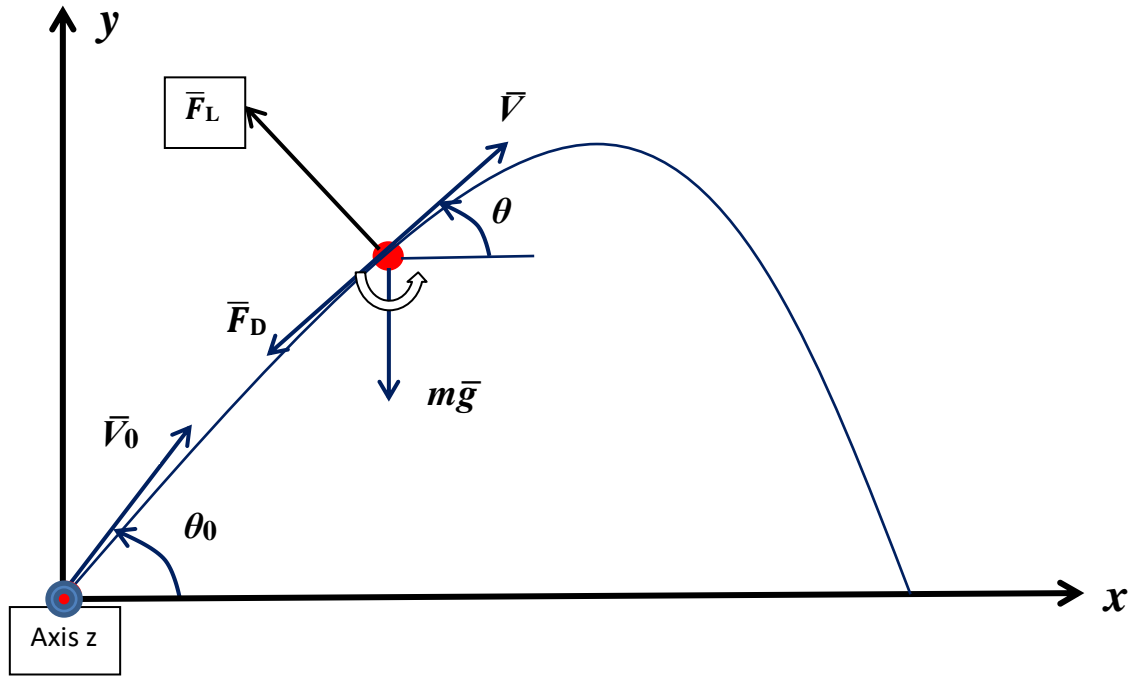

**Figure 1.** Forces acting on the projectile.

## 2. Equations of motion of the projectile

It is now well known that the lift or Magnus force occurs in the perpendicular direction to the axis of rotation and path of the projectile (up or down, relying upon counterclockwise (under-spin)/clockwise rotation (top-spin)). The drag force is tangential to the flight path acting in the opposite direction to the motion. Let us write down the equations of motion of the projectile in projections on Cartesian axes and on natural axes [1,2]:

$$m\frac{dV_x}{dt} = -mgk_1 V V_x \mp mgk_2 V V_y, \quad \frac{dV_y}{dt} = -g - gk_1 V V_y \pm gk_2 V V_x. \tag{2}$$

$$m\frac{dV}{dt} = -mg\sin\theta - mgk_1 V^2, \quad mV\frac{d\theta}{dt} = -mg\cos\theta \pm mgk_2 V^2. \tag{3}$$

The upper signs in the equations correspond to the direction of the Magnus force shown in Figure 1, the lower signs to the opposite direction. In equations (2) the relations are fulfilled: $V_x = V\cos\theta, \ V_y = V\sin\theta.$

The initial conditions are as follows:

at $\quad t = t_0 = 0 \quad$ are fulfilled $\quad x_0 = y_0 = 0, \ V = V_0, \ \theta = \theta_0.$

Equations (2) - (3) have no exact analytical solutions except for special cases [2], [5]. Therefore, almost all the literature deals with the numerical simulation of the equations (2) – (3) by different numerical integration schemes.

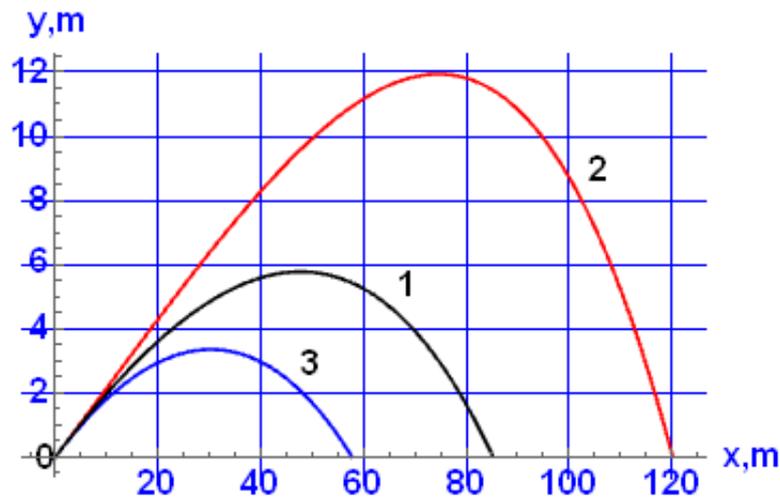

**Figure 2.** Influence of Magnus force on the shape of the projectile trajectory at $y \geq 0$.

**3. Preliminary considerations**

The nature of the simultaneous effect of the medium resistance forces and Magnus forces on the projectile has been studied long ago and quite well when the projectile moves on a limited time interval (from the point of throwing to the point of impact). At the same time, the characteristics of the asymptotic motion of the projectile are of some interest. These characteristics are the ultimate angle of inclination of the projectile trajectory and its terminal velocity. Determination of these parameters is one of the objectives of this study.

Fig.2 shows the trajectories of the golf ball under initial conditions and parameter values from Fig.11 of [1] . The trajectories are calculated for the values $y \geq 0.$ The following values of the parameters included in the formulas (1) were used in the calculations:

$\rho = 1.22 \text{ kg/m}^3, \quad A = \pi r^2 = \pi \times 0.0213^2 \text{ m}^2, \quad \omega = \pm 300 \text{ rad/s}, \quad C_M = 0.167, \quad C_D = 0.45,$

$$m = 0.045 \text{ kg}, \quad g = 9.81 \text{ m/s}^2, \quad k_1 = 0.00088 \text{ s}^2/\text{m}^2, \quad k_2 = 0.00033 \text{ s}^2/\text{m}^2,$$

$$V_0 = 60 \text{ m/s}, \quad \theta_0 = 12°. \tag{4}$$

Curve 1 is plotted at the value of $k_2 = 0$, i.e. in the absence of the Magnus force, curve 2 is plotted at a value of $\omega = -300$ rad/s (reverse spin), curve 3 - at $\omega = +300$ rad/s (straight ball spin). The presence of Magnus forces raises or lowers the trajectory of the projectile and increases or decreases the range.

Figure 3 presents the asymptotic behaviour of the projectile trajectories under at the simultaneous action of the drag force and the Magnus force. Fig. 3 shows the same trajectories as in Fig. 2, but extended into the region of values $y < 0$. The Figure 3 shows that the Magnus forces transform the rectilinear trajectory 1 with a vertical asymptote into inclined straight lines with corresponding asymptotes. When the projectile moves with quadratic resistance and without the Magnus force taken into account, then in the limit the projectile moves along a rectilinear trajectory with known parameters of the trajectory inclination angle $\theta_{term}$ and terminal velocity $V_{term}$ : $\theta_{term} = -90°, V_{term} = 1/\sqrt{k_1}$. The objective of this study is to determine the characteristics of the asymptotic behaviour of projectile trajectories when the Magnus force is taken into account, namely the limit angle of inclination of the trajectory $\theta_{term}$ and terminal velocity $V_{term}$.

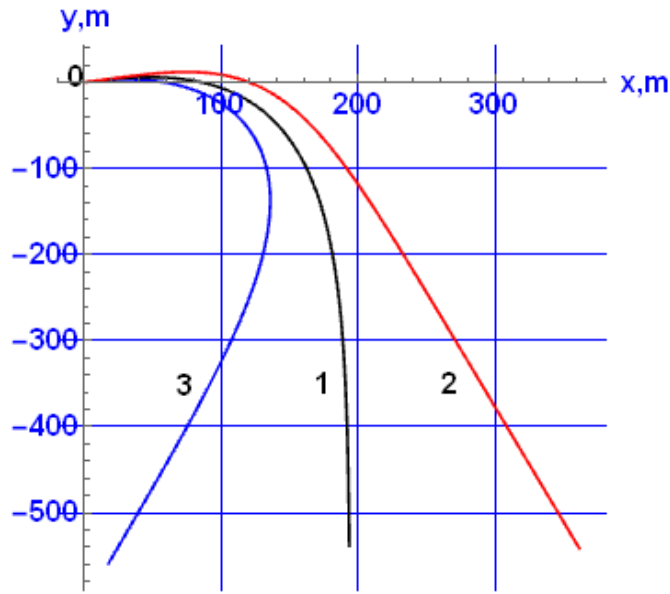

**Figure 3.** Influence of Magnus force on the shape of the projectile trajectory at $y \leq 0$.

To solve this problem, let us pay attention to the behaviour of trajectories 2 and 3 in Fig. 3. Starting from a certain point in time, the projectile moves with a practically constant velocity and a constant angle of inclination of the trajectory to the *x*-axis. In other words, the characteristics of motion $V$, $V_x$, $V_y$, $\theta$ remain constant. This fact can be used to find $\theta_{term}, V_{term}$.

## 4. Determination of asymptotic motion characteristics and velocity hodograph

Let us use equations (2) - (3), assuming backspin of the golf ball ($\omega < 0$). In equations (2) to (3), in this case, the upper signs in the corresponding summands are taken into account. First, let us find the limit angle of inclination of the projectile trajectory to the horizontal. For this purpose, we use the first equation of the system (2) in the form of:

$$m\frac{dV_x}{dt} = -mgk_1 V^2 \cos\theta - mgk_2 V^2 \sin\theta. \tag{5}$$

Let's take into account that starting from some moment of time it is possible to take $V_x = const$. Then in equation (5)

$$k_2 \sin\theta + k_1 \cos\theta = 0. \tag{6}$$

Hence we find the limiting angle of inclination of the projectile trajectory $\theta_{term}$:

$$\theta_{term} = \arctan\left(-\frac{k_1}{k_2}\right). \tag{7}$$

Now, knowing the limiting angle of inclination of the trajectory, we can find the terminal velocity of the projectile. To determine the terminal velocity, we use the second of equations (2). Let us rewrite it in the form

$$\frac{dV_y}{dt} = -g - gk_1 V^2 \sin\theta + gk_2 V^2 \cos\theta.$$

Taking into account that starting from some point in time we can take $V_y = const$, we obtain the equation

$$-g - gk_1 V^2 \sin\theta + gk_2 V^2 \cos\theta = 0.$$

From here we find the terminal velocity of the projectile

$$V_{term} = \frac{1}{\sqrt{k_2 \cos\theta_{term} - k_1 \sin\theta_{term}}} = \frac{1}{\sqrt[4]{k_1^2 + k_2^2}}. \tag{8}$$

At $k_2 = 0$, i.e. at absence of the Magnus force, relations (8) – (9) transform to the known relations $\theta_{term} = -90°, V_{term} = 1/\sqrt{k_1}$. Similar reasoning can be done when investigating projectile motion with a straight ball backspin ($\omega > 0$). In equations (2) – (3) in this case, the lower signs of the corresponding summands should be taken. The results of the study are as follows: $\theta_{term} = \arctan\left(\frac{k_1}{k_2}\right)$, formula (9) does not change.

The second objective of this study is to find the projectile velocity hodograph. To obtain the equation of the velocity hodograph, we use the system of projectile motion equations (3). Dividing the first equation by the second equation, we obtain

$$\frac{dV}{d\theta} = \frac{\left(\sin\theta + k_1 V^2\right)V}{\cos\theta - k_2 V^2}.$$

This expression can be represented in the following form

$$d(V\cos\theta) - \frac{k_2}{3}d\left(V^3\right) - k_1 V^3 d\theta = 0.$$

Let us introduce the notation $V\cos\theta = u$, whence $V = u/\cos\theta$. Then the equation is rewritten in the form

$$du - \frac{k_2}{3}d\left(\frac{u^3}{\cos^3\theta}\right) - k_1 \frac{u^3}{\cos^3\theta}d\theta = 0.$$

Let us transform this relation to the form

$$du - \frac{k_2}{3}\left(\frac{3u^2 du}{\cos^3\theta} + u^3 d\left(\frac{1}{\cos^3\theta}\right)\right) - k_1 \frac{u^3}{\cos^3\theta}d\theta = 0. \tag{9}$$

Multiplier $1/\cos^3\theta$ in the second summand of equation (9) will be considered constant:

$$\frac{1}{\cos^3\theta} = \frac{1}{\cos^3\theta_0} = const. \tag{10}$$

Then equation (9) will take the form allowing its integration

$$du - k_2\left(\frac{1}{\cos^3\theta_0}u^2 du + \frac{u^3}{3}d\left(\frac{1}{\cos^3\theta}\right)\right) - k_1 \frac{u^3}{\cos^3\theta}d\theta = 0.$$

Now let's divide the equation by $u^3$ and integrate. The result of integration is an implicit function $f(V,\theta) = 0$, relating projectile velocity to its trajectory angle:

$$\frac{1}{V^2 \cos^2\theta} + 2k_2\left(\frac{\ln(V\cos\theta)}{\cos^3\theta_0} + \frac{1}{3\cos^3\theta}\right) + k_1\left(\frac{\sin\theta}{\cos^2\theta} + \ln\tan\left(\frac{\theta}{2} + \frac{\pi}{4}\right)\right) - A = 0. \tag{11}$$

Here

$$A = \frac{1}{V_0^2 \cos^2\theta_0} + 2k_2\left(\frac{\ln(V_0 \cos\theta_0)}{\cos^3\theta_0} + \frac{1}{3\cos^3\theta_0}\right) + k_1\left(\frac{\sin\theta_0}{\cos^2\theta_0} + \ln\tan\left(\frac{\theta_0}{2} + \frac{\pi}{4}\right)\right).$$

Under the condition $k_2 = 0$ function (11) transforms into the known [2, 6] equation of the velocity hodograph when the projectile moves with quadratic resistance:

$$V(\theta) = \frac{V_0 \cos\theta_0}{\cos\theta \sqrt{1 + k_1 V_0^2 \cos^2\theta_0 \left(f(\theta_0) - f(\theta)\right)}}, \quad f(\theta) = \frac{\sin\theta}{\cos^2\theta} + \ln\tan\left(\frac{\theta}{2} + \frac{\pi}{4}\right).$$

Under the conditions $k_1 = k_2 = 0$ we obtain, in turn, the equation of the velocity hodograph at parabolic motion

$$V = \frac{V_0 \cos\theta_0}{\cos\theta}.$$

In spite of the used assumption (10), function (11) defines with sufficiently high accuracy the dependence $V(\theta)$ in a wide region of values of motion parameters and initial conditions.

In [2], for the case of relatively small impacts (i.e. $k_1 k_2 \ll 1$) an approximate explicit expression for the velocity hodograph is obtained. In the notations of this paper, it looks as follows:

$$V(\theta) = \frac{V_0 \cos\theta_0}{\cos\theta \sqrt{\frac{2k_2 V_0^2}{3}\left(\frac{\cos^2\theta_0}{\cos^3\theta} - \frac{1}{\cos\theta_0}\right) + 1 + k_1 V_0^2 \cos^2\theta_0 \left(f(\theta_0) - f(\theta)\right)}}. \quad (12)$$

In the next section of the paper, we will compare the accuracy of formulas (11) and (12).

## 5. Results and conclusions

At selected values of design parameters (4) for trajectory 2 from Figure 3, formulas (7) and (8) give values of $\theta_{term} = -69.4°$, $V_{term} = 32.62$ m/s; for trajectory 3 - $\theta_{term} = 69.4°$, $V_{term} = 32.62$ m/s. Numerical calculations based on numerical integration of the system of differential equations of motion (2) by the Runge-Kutta method of the 4th order give exactly the same results.

To check the validity of formula (11), we perform a number of calculations. In making these calculations, we will use a practically applicable range of parameters when playing golf. Assume that the ball is rotating anti-clockwise $(\omega < 0)$. The used parameter ranges are as follows:

$$0 < V_0 \le 75 \text{ m/s}, \quad \left(\text{in calculations } V_0 = 25, 50, 60, 75 \text{ m/s}\right).$$

$$0 < \theta_0 \le 20°, \quad \left(\text{in calculations } \theta_0 = 10°, 15°, 20°\right).$$

$$0 < \omega \le 220\pi \, (691 \text{ rad/s}), \quad \left(\text{in calculations } \omega = 100\pi, 160\pi, 220\pi \text{ rad/s}\right).$$

According to the value $\omega$, coefficient $k_2$ in the calculations takes the values

$$k_2 = 0.00034, \quad 0.000448, \quad 0.000515 \quad s^2/m^2.$$

Coefficient $k_1$ is the same as in the values (4). All figures (except Fig. 7) correspond to the time intervals from the moment when the projectile is thrown to the moment it falls. The throwing and falling points lie on the same horizontal line $y = 0$. The solid blue lines are plotted using formula

(11), the red dotted lines are obtained by numerical integration of the system of differential equations (2) using the Runge-Kutta method of the 4th order.

Figures 4 – 6 are plotted at the following values of the projectile parameters.

Figure 4. Dependence $V = V(\theta)$ for $V_0 = 60$ m/s, $k_2 = 0.000515$ s$^2$/m$^2$, $\theta_0 = 10°, 15°, 20°$. Curve 1 is plotted for the value $\theta_0 = 20°$.

Figure 5. Dependence $V = V(\theta)$ for $\theta_0 = 15°$, $k_2 = 0.000448$ s$^2$/m$^2$, $V_0 = 25, 50, 75$ m/s.

Figure 6. Dependence $V = V(\theta)$ for $\theta_0 = 15°$, $V_0 = 60$ m/s, $k_2 = 0.00034, 0.000448, 0.000515$ s$^2$/m$^2$.

Figs. 4– 6 demonstrate high accuracy of formula (11) in the above ranges of parameters $V_0, \theta_0, \omega$.

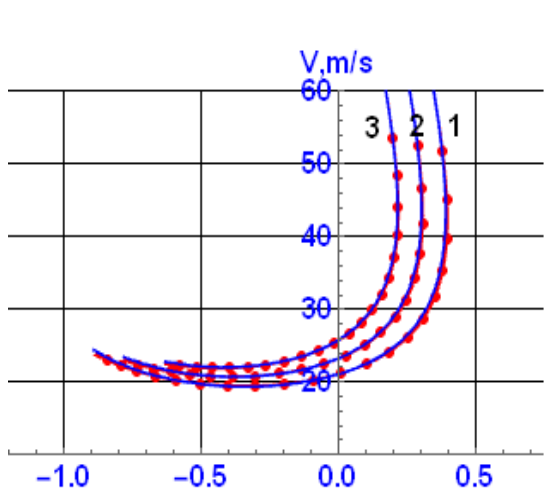

Figure 4.

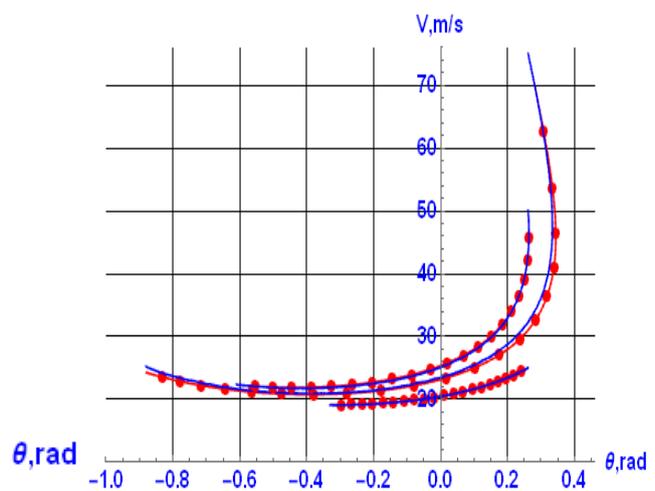

Figure 5.

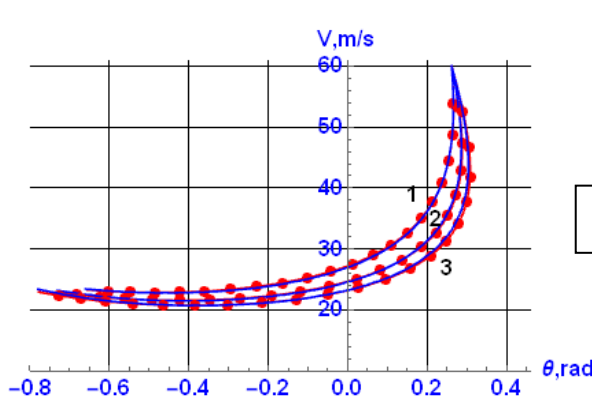

Figure 6.

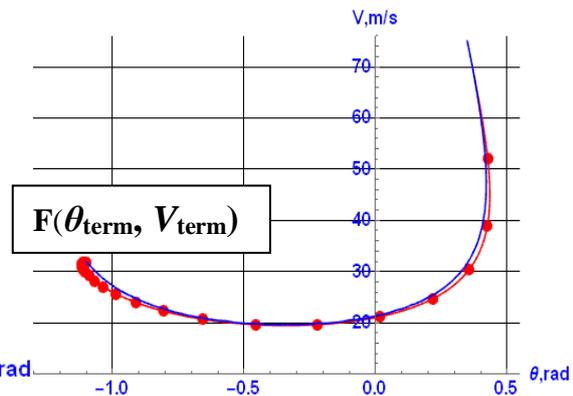

Figure 7.

Figure 7. Dependence $V = V(\theta)$ for $\theta_0 = 20°$, $V_0 = 75$ m/s, $k_2 = 0.000448$ s$^2$/m$^2$.

In Fig. 7 the dependence $V(\theta)$ is plotted for asymptotic projectile trajectory. The time of motion in this case can be as large as desired. Curve $V(\theta)$ in this case is bounded by the final point. This final point in Fig. 7 is denoted as $F$ and has coordinates $F(\theta_{term}, V_{term})$, according to formulas (7)– (8). Here $\theta_{term} = -63.0° = -1.1$ rad, $V_{term} = 31.8$ m/s. Figure 7 shows that for asymptotic projectile trajectories as well, formula (11) provides remarkable accuracy.

It is of interest to compare the accuracy of formulas (11) and (12) at the values of parameters $V_0, \theta_0, \omega$ used in the paper. Fig. 8 shows the dependences $V = V(\theta)$ for

$$\theta_0 = 15°, \quad V_0 = 60 \text{ m/s}, \quad k_2 = 0.000448 \quad \text{s}^2/\text{m}^2. \tag{13}$$

The blue curve is constructed by formula (11), the black curve is constructed by formula (12). The red point curve is obtained by numerical integration of the system of equations (2). Figure 8 shows that formula (11) approximates the projectile velocity hodograph very well for parameter values (13), unlike formula (12). However, if we reduce the value of the coefficient $k_2$ by a factor of ten for the same values of the parameters $V_0, \theta_0$, the curves given by formulae (11) and (12) almost coincide. This fact is illustrated in Figure 9. Thus, formula (12) is workable at very small values of parameters, while formula (11) defines the velocity hodograph in a much wider range of parameters.

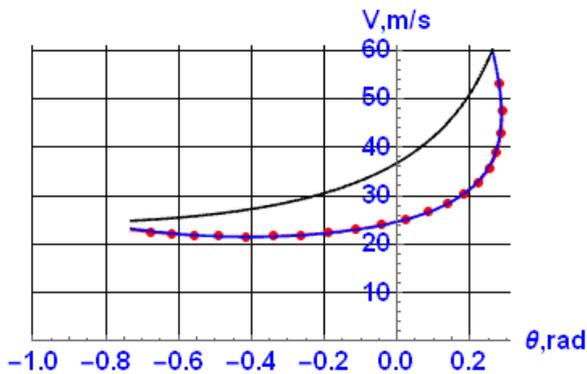
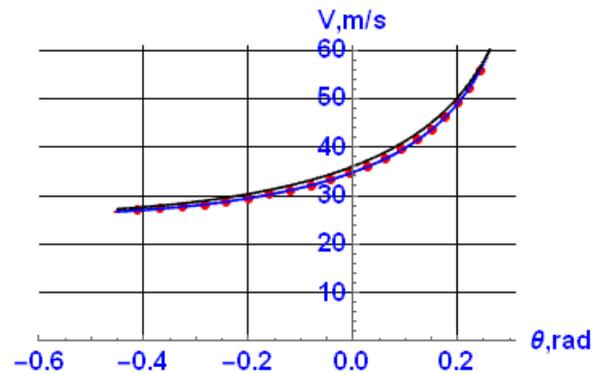

**Figure 8**.   **Figure 9**.

Note that formula (11) is obtained under the assumption of the reverse rotation of the projectile $(\omega < 0)$. In the case of direct rotation $(\omega > 0)$ it is necessary to change the sign of the coefficient $k_2$ in formula (11). The case of forward rotation in the asymptotic motion of the projectile requires further study.

Thus, based on the general character of the asymptotic motion of the projectile (i.e., uniform motion along the asymptotes) and the equations of motion themselves, we can determine the limiting trajectory angle and terminal velocity. This does not require numerical integration of the differential equations of motion of the projectile. In addition, an approximate high-precision formula relating the trajectory angle of the projectile to its velocity is derived. The presented formula for the velocity hodograph can be employed to estimate and optimize the kinematic characteristics of the projectile. It is shown that the formula has high accuracy in the used range of projectile parameters. In general, the derived formulae can be useful to all researchers of projectile motion.